\documentclass{aa-multiplecitationsrepair}

\usepackage{graphicx}
\usepackage{txfonts,hyperref}

\usepackage{color,xcolor,xspace,pdflscape,dblfloatfix}

\newcommand{\mb}[1]{ #1}
\definecolor{mhi}{rgb}{0.6,0.0,0.6}

\newcommand{\equ}[1]{Eq.~\ref{eq:#1}}

\newcommand{\fig}[1]{Fig.~\ref{fig:#1}}

\newcommand{\tab}[1]{Table~\ref{tab:#1}}

\newcommand{\sect}[1]{Sect.~\ref{sec:#1}}
\newcommand{\app}[1]{Appendix~\ref{app:#1}}

\newcommand{\abar}[0]{\ensuremath{a_\mathrm{bar}}\xspace}
\newcommand{\aobs}[0]{\ensuremath{a_\mathrm{obs}}\xspace}
\newcommand{\alsbs}[0]{\ensuremath{a_\mathrm{LSBS}}\xspace}
\newcommand{\rh}[0]{\ensuremath{r_{1/2}}\xspace}
\newcommand{\re}[0]{\ensuremath{R_\mathrm{e}}\xspace}

\usepackage{array}
\usepackage{makecell}
\usepackage{ragged2e}

\let\oldtimes\times
\renewcommand{\times}{\mathbin{\!\oldtimes\!}}

\begin{document}

\title{A correlation predicting galaxies without dark matter}

\authorrunning{Michal B\'ilek}

   \author{
   Michal B\'ilek \inst{1}
   }

   \institute{$^1$ Astronomical Observatory of Belgrade, Volgina 7, 11060 Belgrade, Serbia,     \email{michal.bilek@aob.rs}
             }
   \date{Received: ...; accepted ...}

 \abstract{\mb{The standard theory of galaxy formation predicts that all galaxies should contain dark matter, yet a handful of recently discovered galaxies appear to lack it, challenging our understanding of galaxy formation. We investigate whether such dark-matter deficient objects can be identified from their baryonic properties alone, analogously to the radial-acceleration relation, which tightly links baryon and dark matter distributions in spiral galaxies.  Using a sample of ultra-diffuse and dwarf spheroidal galaxies---systems whose baryonic properties resemble those of the confirmed dark-matter-deficient galaxies---we systematically search for a formula to predict baryonic fractions from stellar mass, effective radius, distance to the host, and the host's baryonic mass. We find that baryonic fraction correlates most strongly with the gravitational acceleration expected from baryons alone, $\abar$, or equivalently, with mean surface brightness, following an approximately $\abar^{-1}$ dependence. This scaling resembles the radial-acceleration relation but differs in functional form and applies to a different galaxy population. Strikingly, the dark-matter-deficient galaxies occupy the extreme end of the correlation. This suggests that they result from standard formation processes operating at unusual intensities rather than from exotic mechanisms. Importantly, the correlation predicts that all ultra-diffuse galaxies brighter than approximately 25\,mag\,arcsec$^{-2}$ in the $g$-band should have very low dark matter content, offering a straightforward observational criterion for identifying these rare objects.}
 }

   \keywords{ Galaxies: dwarf --- Galaxies: Local Group --- Galaxies: kinematics and dynamics    --- Gravitation --- Galaxies: evolution --- Galaxies: structure }
               
   \maketitle
\section{Introduction}

\mb{According to the current mainstream theories, galaxies are embedded into dark matter halos \citep{mo}. The mass of the halo can exceed the baryonic mass of the galaxy (that means the gas and stars) hundred or several thousand times \citep{wechsler18}. The contribution of the dark matter to the gravitational field of a galaxy exceeds that of the baryons many times, except potentially in the galaxy centers \citep{behroozi13}. {Without dark matter, galaxies would not have time to reach their observed baryonic masses within the age of the Universe \citep{peebles80}}. Therefore it came as a big surprise when a galaxy lacking dark matter, called NGC\,1052-DF\,2 (or just DF\,2), was discovered \citep{vandokkum18}. It belongs to the class of ultra-diffuse galaxies (UDGs,\citealp{vandokkum15}), which is galaxies with { baryonic  masses of  dwarfs but much larger effective radii}. It has a spheroidal shape, {negligible rotation \citep{emsellem19}}, and does not contain gas. The appearance resembles the dwarf-spheroidal galaxies in the vicinity of our galaxy, the Milky Way (MW), but is larger and more massive. {Up to now, we know only three more spheroidal dark-matter deficient galaxies: NGC\,1052-DF\,4 (or DF\,4) \citep{shen23}, FCC\,224 \citep{buzzo25}, and NGC\,1052-DF\,9 (or DF\,9) identified very recently \citep{keim26}. The objects DF\,2, DF\,4 and DF\,9 seem to members of the NGC\,1052 galaxy group.} Some rotating gas-rich UDGs were reported to be lacking dark matter too \citep{Mancera-Pina2019,chen24}, even though this might be because of systematic errors \citep{banik22b,lelli24}. Hereafter, we will consider only the spheroidal dark-matter deficient galaxies which are mostly pressure supported \citep{chilingarian19}. Various formation scenarios of these objects have been proposed \citep[e.g.,][]{ogiya18,silk19,muller19b,yang20b,jackson21,trujillogomez22,zhang25}, {including that the low dark matter content is an illusion stemming from systematic errors} \citep{trujillo19,montes20,montes21,Shen2021,beasley25}. According to the currently leading {scenario, dark matter deficient galaxies} form in rare high-speed collisions of gas-rich dwarfs in the early Universe \citep{vandokkum22, keim25}. The collision leads to a segregation of dark and baryonic matter from the otherwise normal progenitors. In general, UDGs resemble dwarf spheroidals in many regards. Their properties seem to form a single continuum,  without any obvious sign of bimodality in any property \citep{conselice18,ruizlara18,marleau21,ferremateu23,zoller24,buzzo24}. In this work, we treat dwarf spheroidal and ultra-diffuse galaxies as a single class of objects and call them low-surface-brightness spheroids (LSBSs).}

 \begin{figure*}
\centering
   \includegraphics[width=17cm]{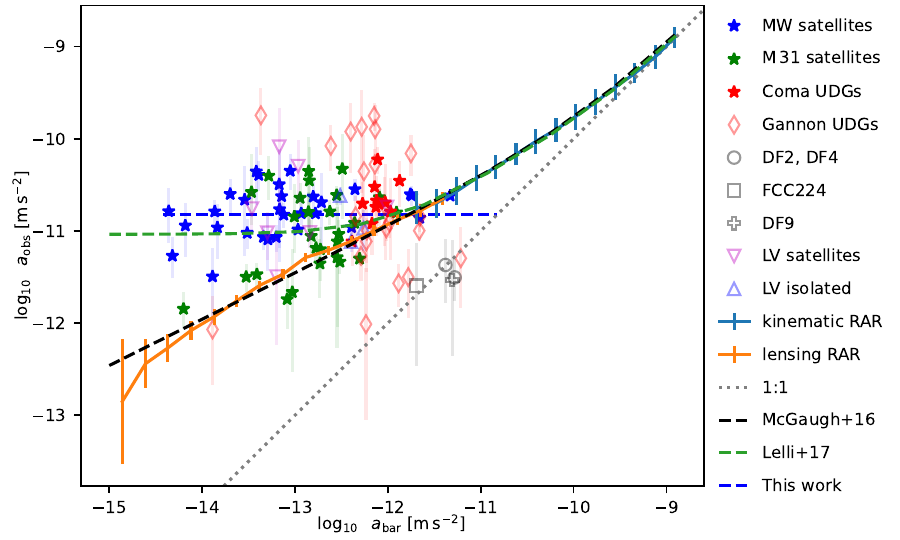}
     \caption{Dwarf spheroidal and ultra-diffuse galaxies do not follow the RAR. The points are all the galaxies investigated in this paper coming from recent sources (see \app{data}).}
     \label{fig:rar}
\end{figure*}
 
\mb{The question of the formation of the dark-matter deficient galaxies is a part of a more general puzzle. The radial acceleration relation (RAR) links the gravitational acceleration expected from the distribution of baryons and the Newton gravitational law $a_\mathrm{bar}$  (that is the gravitational force felt by one kilogram of matter) and the actually observed gravitational acceleration $a_\mathrm{obs}$  (resulting as the sum of the contributions by the baryons and dark matter). Its existence was first proposed on theoretical grounds by the Modified Newtonian Dynamics (MOND, \citealp{milg83a}), a hypothesis of modification of the law of gravity or  inertia. After the initial success in the form of the tests of MOND (see \citealp{famaey12} for a review), the RAR was established as an empirical scaling relation by \citet{mcgaugh16} (the blue line in \fig{rar}) where it emerged from the analysis of a large sample of rotation curves of spiral galaxies with precise baryonic mass models. The fit to this data (the dashed black line in \fig{rar}), matches the MOND prediction, which says that the RAR consists of two power-law regimes separated at the acceleration of $a_0=1.2\times 10^{-10}\,$m$\,$s$^{-2}$.  The RAR in spirals is extremely tight \citep{li18,stiskalek23,desmond23,varasteanu25}. The RAR allows predicting the rotation curves of spiral galaxies.  The existence of this correlation implies that dark and visible matter must be organized such that the RAR holds true at any galactocentric radius. In the context of the dark matter hypothesis, the origin of the RAR is not fully understood \citep{milg02b,wu15,dicintio16,milg16,keller17,ludlow17,tenneti18,brouwer21,li22} which gave rise to proposals to assign the dark matter particles unusual properties to guarantee the validity of the RAR \citep{berezhiani15,berezhiani18,blanchet08,blanchet09,famaey20}. \citet{lelli17} found that giant elliptical and lenticular galaxies follow the same relation. However, dwarf spheroidal satellites of the MW and Andromeda  (M\,31) galaxy generally showed greater observed accelerations than predicted by the formula by \citet{mcgaugh16} and MOND, such that the correlation flattened out for low $a_\mathrm{bar}$ (the green and blue stars in \fig{rar} show the latest measurements). \citet{lelli17} warned that disagreement could possibly be because of systematic errors, such as the satellites not being in dynamical equilibrium. That time, only the dwarf spheroidal satellites were available to probe the lowest acceleration regime, and thus \citet{lelli17} fitted the empirical relation by the curve shown in \fig{rar} by the green dashed line. Later, the analysis of gravitational lensing around giant spirals and ellipticals showed (the orange line in \fig{rar}, \citealp{mistele24,mistele24b}) that their gravitational field closely follow  the formula found by \citet{mcgaugh16} even at the weakest accelerations. That was independently   confirmed by the motions of galaxies in groups \citep{milg18,milg19}. This means that the dwarf spheroidals investigated by \citet{lelli17} do not lie, for whatever reason, on the RAR formed by the other galaxies.

By now, substantially more kinematic measurements of LSBSs have been done since the times of \citet{lelli17}  (an extensive compilation is shown as the data points in \fig{rar}). Some of them are satellites of the nearby galaxies, others not and some are UDGs, some of which are dark-matter deficient (the galaxies advertised in the literature as dark-matter deficient are shown by the black symbols in  \fig{rar}). The UDGs have larger sizes and masses than typical dwarf spheriodals and we measure their velocity dispersion from the integrated light rather than from individual stars. One would then expect that different LSBSs would not be affected by the same systematic errors. Yet, in the $\aobs$-$\abar$ space, all LSBSs  seem to form a common cloud of points (\fig{rar}). It seems to form a sequence different from the RAR of spirals and giant ellipticals. The dark-matter deficient galaxies seem to be members of this alternative sequence.


In this paper, we assume that the measurements of velocity dispersion of LSBSs are indeed not affected by systematic errors, that is that the LSBSs do not follow the RAR of spirals inherently.  Given that the baryonic structural properties of galaxies (such as luminosity and effective radius) follow different sequences for different galaxy types  \citep{kormendy09,kormendy12, hua26}, it would not be too surprising if also their dark matter fractions followed different sequences. Some observations hint that galaxy clusters \citep{the88,sanders94,sanders99,sanders03,milgcbdm,angus08,ettori19,li24,kelleher24,famaey25, mistele25} and their central giant elliptical galaxies \citep{richtler08,hilker18,bil19,tian24,bil26} do not follow the RAR either.

The main purpose of this work is to systematically search for an analog of the RAR for LSBSs, in the sense of a formula that would allow predicting their gravitational fields based on the distribution of baryons. Specifically, we aim to find a formula to predict the baryonic fractions of LSBSs on the basis of their stellar masses, effective radii, distances from the hosts and the masses of the hosts. The formula should not only match the data, but also satisfy certain general theoretical requirements. The method was motivated as a search for a MOND-like theory that would work for LSBSs, but the result is a fully empirical finding. The identified correlation allows estimating the dark matter fractions of LSBSs  just from their appearance in images and the known distance from the Earth. Perhaps its most striking consequence is that it allows predicting which LSBSs will be dark-matter deficient. This opens the possibility to detect these as-so-far rare galaxies in large quantities. The correlation also suggests that the dark-matter deficient galaxies form by standard processes acting at unusual intensities rather than by qualitatively rare mechanisms. Finally, whatever mechanism gives rise to the RAR in the giant galaxies, it seems to have another mode that operates in the LSBSs.}

\begin{figure}
    \centering
    \includegraphics[width=1\linewidth]{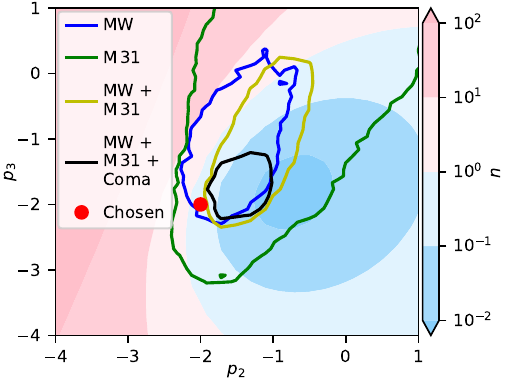}
    \caption{Choosing the parameters  $p_2$ and $p_3$ in the ansatz of  \equ{xansatz}. The contour outlines indicate the regions preferred respectively by the satellites of the Milky Way, Andromeda galaxy, { the union of all these satellites, and the union of all satellites with} the ten UDGs in the Coma cluster. The filled contours indicate the value of the normalization constant when fitting all data by the power law in \equ{norm}. The red point indicates the adopted values of $p_2$ and $p_3$ that satisfy our criteria the best.}
    \label{fig:normmap}
\end{figure}

\section{Finding the correlation}
\label{sec:finding}
\mb{We aim to find a correlation to predict the baryonic fractions of LSBSs based only on their baryonic properties. We first do that using} only satellites of the MW  and M\,31 galaxies and the UDGs in the Coma Cluster. These are the objects for which we know all necessary parameters the most precisely. Once the relation is found, we show that the objects that were not included in its search follow it too. Details on the galaxy samples and all data can be found in \app{data}. 

Due to observational difficulties, in LSBSs it is usually possible to measure the gravitational fields only at the 3-D deprojected half-mass radius \rh. The formula of \citet{wolf10} is used to get the dynamical mass $m_\mathrm{D,1/2}$ (the sum of the baryonic and dark mass) cumulated within \rh from the mean line-of-sight velocity dispersion of the galaxy $\sigma$ as $m_\mathrm{D,1/2} = 3 G^{-1}\sigma^2 \rh$. The radius \rh can be estimated from the observed effective radius \re as  $\rh\approx \frac{4}{3} \re$ \citep{wolf10}. If we combine this with Newton's gravitational law, we get that the gravitational acceleration at \rh is:
\begin{equation}
\aobs = \frac{9}{4}\frac{\sigma^2}{\re}.    
\label{eq:aobs}
\end{equation}
This formula assumes that the galaxy does not deviate much from a non-rotating sphere in the dynamical equilibrium. The acceleration due to the baryons in a galaxy of a total baryonic mass $m$ at \rh is:
\begin{equation}
\abar =  \frac{G\,0.5m}{\left(\frac{4}{3} \re\right)^2}.
\label{eq:abar}
\end{equation}

We seek a formula that would allow predicting the ratio 
\begin{equation}
y = \frac{\aobs}{\abar},    
\end{equation}
which has the meaning of the ratio of the dynamical and baryonic mass cumulated below \rh, that is $m_\mathrm{D,1/2} /m_{1/2}$ where $m_{1/2}=0.5m$. We allow $y$ to be a function of \re, $m$, and $G$. Since the successful RAR involves $a_0$, we allow our function to involve $a_0$ too. Next, we allow a dependency on the distance $D$ from the host and the mass of the host $M$, since the object might be affected, for example, by tidal forces. In total, we seek a function:
\begin{equation}
y = f(m, M,\re, D, G, a_0).    
\end{equation}
We require the formula to be dimensionally correct. This implies that the function $f$ must have the form of:
\begin{equation}
y=f(\phi_1,\phi_3,\phi_3), 
\end{equation}
where
\begin{equation}
\phi_1=\frac{m_{1/2}}{M},~~~~ \phi_2=\frac{\rh}{D},~~~~ \phi_3=\frac{D}{\sqrt{GM/a_0}}.    
\end{equation}
For simplicity, we restrict $f$ to have the form of:
\begin{equation}
y = f(x),    
\end{equation}
where $f$ is any monotonic function,
\begin{equation}
x=\phi_1 ~ \phi_2^{p_2} ~ \phi_3^{p_3},
\label{eq:xansatz}
\end{equation}
and $p_2$ and $p_3$ are constants. We note that adding a power to $\phi_1$ is not necessary, since this is equivalent to a suitable choice of $f$.

\begin{figure}
    \centering
    \includegraphics[width=1\linewidth]{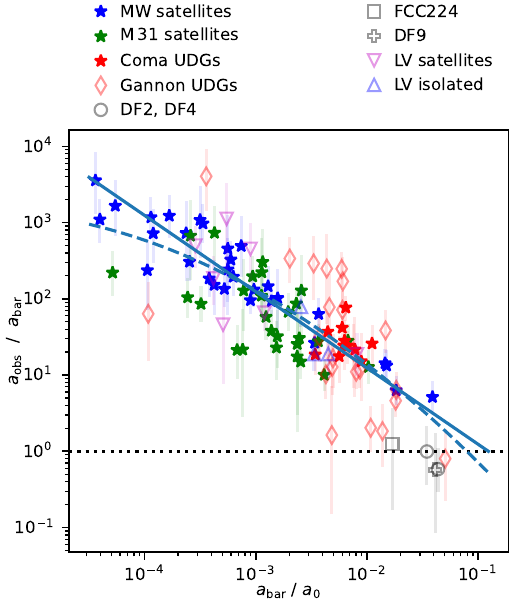}
    \caption{{The {reported} correlation between the baryonic fraction and baryonic acceleration.} The filled data points were used for finding the correlation. The empty points were used for verifying that the correlation applies to all LSBSs. The black symbols mark the galaxies that were reported in the literature as dark-matter deficient: the galaxies DF\,2, DF\,4 and FCC\,224. The dotted line indicates no mass discrepancy, that is no need for dark matter within \rh. \mb{The blue full line is the regression  by \equ{result}, the blue dashed line by \equ{poly}}.}
    \label{fig:all}
\end{figure}

\begin{figure*}[b!]
  \includegraphics[width=17cm]{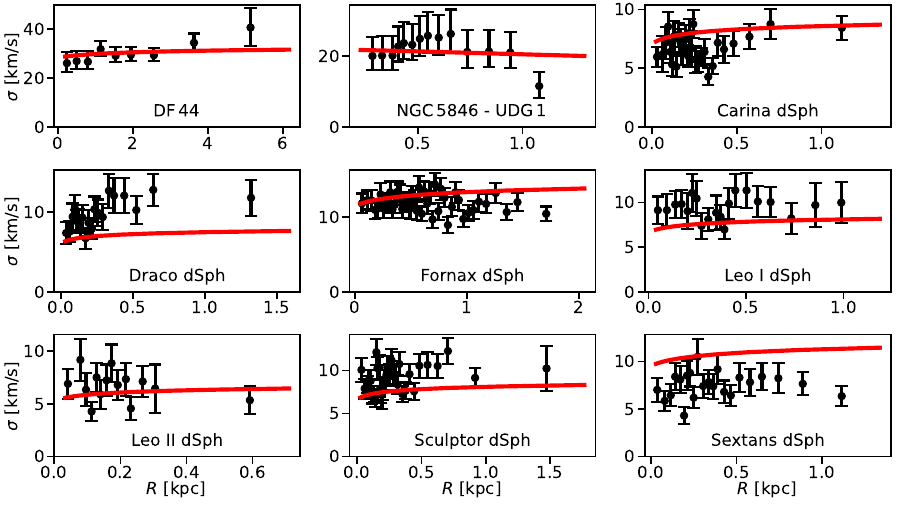}
     \caption{No-fitting models of the observed velocity dispersion profiles of LSBSs based on the force law \equ{const}.}
     \label{fig:resolved}
\end{figure*}

In practice, we calculated for each galaxy $y$, $\phi_1$, $\phi_2$ and $\phi_3$. Then we defined a grid of the values $p_2$ and $p_3$ and for each galaxy we calculated $x$ through \equ{xansatz}. We assigned to each grid point a value that is equal to the Spearman correlation coefficient of the correlation $x$ vs. $y$. The Spearman coefficient quantifies how well the points form a monotonic sequence. We experimented with assigning the points weights according to their measurement uncertainties, but it did not change the results substantially. We then for simplicity continued with the standard Spearman coefficient that ignores uncertainties. In this way, we found which combination of  $p_2$ and $p_3$ gives the strongest correlation. In \fig{normmap}, we indicate by the contour lines the regions that differ from the best correlation by less than two $\sigma$ (see \app{spearmann} for details). Such contours are shown for the full sample, and separately the MW satellites and the M\,31 satellites to demonstrate that all samples give consistent results.

We noted that for the $p_2$ and $p_3$ giving the strongest correlation, the data form a power law
\begin{equation}
y=nx^\alpha,    
\label{eq:norm}
\end{equation}
and therefore we hereafter assumed such a form of $f$. Other than $x$ and $y$ to correlate strongly, we imposed that the powers  $p_2$ and $p_3$  should be small integers or their ratios. We also required the sought for formula not to involve constants much different from unity, so we required $n$ to be close to one. Its map as a function of  $p_2$ and $p_3$ is shown in \fig{normmap} by the full contours.  The choice of $p_2= p_3=-2$ is a good tradeoff of all these criteria. We are then getting:
\begin{equation}
    x~=~\frac{m_{1/2}}{M}~ \left(\frac{\rh}{D}\right)^{-2}~\left(D\sqrt{\frac{a_0}{GM}}\right)^{-2}=~\frac{Gm_{1/2}}{a_0 \rh^2}.
\end{equation}
Interestingly, the variables $D$ and $M$ have canceled out in this relation, such that the mass discrepancy
\begin{equation}
y~=~f\left(\frac{G\,m_{1/2}}{a_0\, \rh^2 }\right)  = f\left(\frac{\abar}{a_0}\right)   
\end{equation}
is a function of $\abar/a_0$   like in the RAR.

Figure~\ref{fig:all} shows the data in the $x$-$y$ space as the full stars. A direct fitting of the data would probably lead to confusing results, because of the systematic effects involved, such as the tidal forces. Instead, we note that the slope $\alpha$ is close to $-1$ and adopt this value. Now, fitting the normalization $n$ in the log-log space in
\begin{equation}
\frac{\aobs}{\abar}  = n \left(\frac{Gm_{1/2}}{a_0\rh^2}\right)^{-1}, 
\label{eq:result}
\end{equation}
taking into account the uncertainty in $y$ (\app{data}), we are getting $n=0.12\pm0.02$. The fit is plotted in \fig{all} by the full line.  The reduced chi-square of 0.16 indicates that the model is perfectly consistent with the data. We note that the right hand side of \equ{result} is proportional to $\abar^{-1}$, meaning that $\aobs$  is a constant that we call \alsbs: 
\begin{equation}
\aobs ~=~ \alsbs ~:=~n\, a_0 ~=~1.5\times 10^{-11}\, \textrm{m}\,\textrm{s}^{-2}.   
\label{eq:const}
\end{equation}


Now, we can test \equ{result} using an independent galaxy sample.  It is formed by the UDG catalog by \citet{gannon24}, the dark-matter deficient galaxies DF\,2, DF\,4\mb{, DF\,9}  and FCC\,224, and the dwarf spheroidal galaxies in the nearby universe that either are or are not satellites of bigger galaxies (see \app{data} for all data details). These data are shown in Fig.~\ref{fig:all}  by the empty symbols. They follow the correlation found before. If we use the original and the new samples together, the fit of $n$ in \equ{result} comes out $n=0.12\pm0.01$, consistently with the main sample. If we use only the additional objects, we get $n=0.13\pm0.03$ and a reduced chi-square of 0.32. The correlation is thus independent of the sample choice. The law \equ{const} is plotted in \fig{rar} by the blue dashed line.

\mb{We consider also a more complex parametrization of the data in \fig{all}, since the correlation appears curved. Fitting a second degree polynomial to all data  yields:
    

\begin{equation}
    \log_{10} y = -0.158(\log_{10}x)^2 -1.76\log_{10}x -1.74.
    \label{eq:poly}
\end{equation}
The Akaike information criterion strongly prefers this model over the one-parametric model in \equ{result} ($\Delta$AICc = 40). Nevertheless, we remind again that this more complex fit might rather pick up the effects systematic errors and therefore we hesitate to draw strong conclusions here.}

\subsection{Test on resolved velocity dispersion profiles}
\label{sec:resolved}
\mb{The results of the previous section were derived for a specific radius, namely \rh, for each galaxy. We have not clarified yet if the ``constant gravity law'' \equ{const}} is valid only at this position, or anywhere within the galaxies.  Here we test that making use of the resolved velocity dispersion profiles that are available for nine of the galaxies. For them, we solved the isotropic spherical Jeans equation \citep{binney-tremaine87,bil19,samurovic19} to obtain the velocity dispersion profiles, assuming that the force law from \equ{const} works at all radii. The results are in \fig{resolved}. Here no fitting was applied; we just used the most probable values of the galaxy parameters from the literature (\app{data}). 

The observed velocity dispersion profiles appear rather flat. If \equ{const} were not valid at all radii, the modeled profiles could be strongly increasing or decreasing and potentially diverging at zero. But this is not happening, they are rather flat too. All models are in the right ballpark and some are already fully satisfactory. For some galaxies, the models would benefit from some mild tuning of the tracer population profile and of the anisotropy parameter and galaxy rotation. There might also be an influence by tides and the non-sphericity of the object. A detailed modeling is however beyond the scope of this first paper on the topic. {This result agrees with the finding of \citet{lelli17}, who found that in Fornax and Sculptor, \aobs agree at two different radii. They used a method that is relatively insensitive to the anisotropy parameter.} We conclude that the constant gravity law of \equ{const} describes the data within the whole observable extents of the galaxies reasonably well. \mb{The polynomial gravity model \equ{poly} gives very similar results.}

\section{Discussion and consequences}
\label{sec:disc}

{Our results align with the findings of \citet{lelli17}, who reported that the observed acceleration in the dwarf spheroidal satellites of the Milky Way and M\,31 shows no correlation with the baryonic acceleration. \mb{Here we have newly showed that this is the optimal parametrization in a certain sense. Actually, the data for the MW and M\,31 are not sufficient}  to establish that \aobs is independent of $m_{1/2}$, \rh, $M$, or $D$, as \fig{normmap} demonstrates. This conclusion requires additional data that became available after the publication of \citet{lelli17}, along with a systematic functional search that we have performed here.}

At the moment being, we do not know what physics \mb{the correlation between the baryonic fraction and \abar}  stems from. For now, it has to be treated as empirical. It is well possible that \equ{result} is rather a well-working regression of data rather than the proper equation that nature follows. The correct function must nevertheless be numerically close to it. {It is also possible that \mb{the correlation} is a result of systematics errors, even though  the fact that it is followed by objects $\sim6$ orders of magnitude in stellar mass seems encouraging.} 

The found correlation resembles the RAR in several regards. Equation~\ref{eq:const} tells us that in LSBSs, the dark and visible matter are spatially organized such that the sum of their contributions to the gravitational field is constant. This resembles the situation with the standard RAR, where the sum of the two contributions is a universal function of the baryonic acceleration. The result can also be expressed in the form of \equ{result}. This can be read as that the mass discrepancy is a function of the baryonic acceleration. This is the same as for the RAR, but the function is different. Finally, the constant \alsbs is not so far from the acceleration scale of the RAR because  $\alsbs/a_0=0.15\pm0.01$. This suggests that whatever mechanism causes the RAR, it has another mode, which gives rise to {the relation \equ{const}}.

An opened question remains of what objects follow the RAR and which \equ{result} -- and which potentially another correlation. The current empirical status stands like this: 1) The LSBSs follow \equ{result}. 2) There are indications that the central giant galaxies of clusters follow the RAR but with a higher acceleration scale \citep{tian24} (just as galaxy clusters as wholes \citealp{tian20,tian24}).  3) The other galaxies follow the standard RAR. Further research is needed to clarify this confusing situation. 

Extrapolating \equ{result} toward $\abar >\alsbs$  would lead to the exotic prediction that the acceleration is lower than expected from the baryons alone. It is not clear what this means. Perhaps no objects in the ``LSBS regime'' exist beyond \alsbs.  Another possibility is that similarly to the RAR, the acceleration saturates at the baryonic value, meaning that the total prescription for the gravitational field reads:
\begin{equation}
\aobs = \begin{cases}
~\abar, & \text{if} ~~~~ \abar\gg\alsbs,\\
~\alsbs, & \text{if} ~~~~ \abar\ll\alsbs.
\end{cases}
\end{equation}
Also, there must be a radius at which the constant gravity law breaks in order to keep the dark matter halo masses finite.

Regardless of these open questions, the found correlation has important consequences. A remarkable example is that it predicts that some galaxies should be dark matter free. Interestingly, the correlation was deduced only from the objects with the normal amount dark matter. Only after that we found that the dark-matter deficient galaxies follow it too. This means that in principle, the existence of dark-matter deficient galaxies could have been predicted before their discovery.  \mb{Actually, the fact that the DF\,9 galaxy is dark-matter deficient has been discovered after the first submission of this paper. The galaxy is located at the bottom-right tip of the could of points formed by all the other LSBSs in \fig{all}, that is where the correlation predicts such galaxies to be. }

\begin{figure}
    \centering
    \includegraphics[width=1\linewidth]{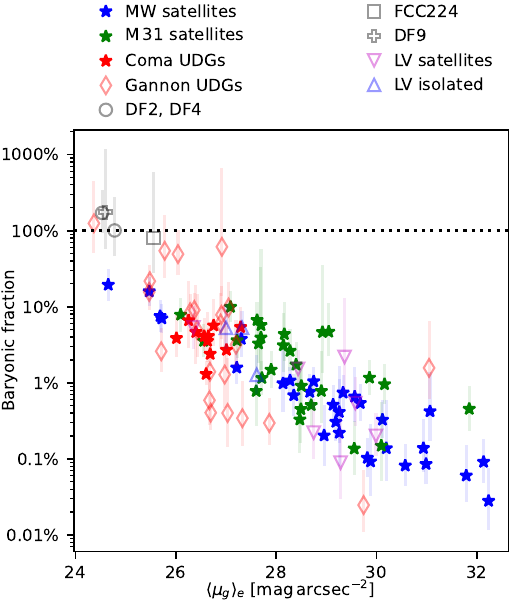}
    \caption{{The reported correlation allows predicting the baryonic fractions of LSBS galaxies just from photometry, namely from the average surface brightness  within one effective radius $\langle \mu_g\rangle_e$. The dotted line indicates no mass discrepancy, that is no need for dark matter within \rh. The meaning of the symbols is as in \fig{all}. A stellar mass-to-light ratio in $g$ of two was assumed for all objects.}}
    \label{fig:sb}
\end{figure}

In particular, the formula predicts that all LSBSs (and perhaps other object types) that have their baryonic accelerations close to $\alsbs$  will be dark matter deficient. This gives us the possibility to discover many new dark-matter deficient galaxies, just on the basis of their images. To give a particularly convenient method for that, we note that the baryonic acceleration at some radius is proportional to the average surface density within that radius. {The correlation from \fig{all} is replotted in \fig{sb} in terms of more observationally friendly parameters, that is the baryonic fraction as a function of the average surface brightness in the effective radius in the $g$ band (assuming the mass-to-light ratio of two).} \mb{The figure shows that the currently known dark-matter deficient galaxies start appearing at the surface brightness of  around 25$\,$mag$\,$arcsec$^{-2}$ in $g$. If we take the fits \equ{result} or \equ{poly} at the face value,  LSBS should have a zero dark matter content at $\abar = 0.1a_0$, which translates to a surface brightness of 23.6$\,$mag$\,$arcsec$^{-2}$ in $g$.} In \fig{sb}, the rightmost red diamond corresponds to the galaxy Hydra-I~UDG\,4. It has not been pointed out in the literature as a dark-matter deficient galaxy, but according to our correlation, it should be even poorer of dark matter than the famous case of DF\,2. {The correlation in \fig{sb} and \fig{all} is a bit curved, which is in line with the above note that the power-law parametrization \equ{result} is just a convenient approximation. More data are necessary before making firm conclusions on this.}

Recent literature has identified the dark-matter deficient galaxies as products of rare events, namely of high-speed collisions of gas-rich dwarf galaxies \citep{vandokkum22,keim25}. Our results seem to contradict this explanation because the dark-matter deficient galaxies lie at the extrapolation of the correlation formed by the galaxies with the standard dark matter content. If the dark-matter deficient galaxies were products of exotic mechanisms such as collisions of gas-rich dwarfs, their horizontal position in \fig{all} could be any, yet they happen to lie at the correlation formed by the galaxies with the standard dark matter content. It rather seems that in the dark-matter deficient galaxies, the standard processes that control the amount of baryons in any galaxy (such as winds from stars, supernova explosions, and active galactic nuclei) acted at an unusual intensity.   

Also, it has been proposed that the dark-matter deficient galaxies only appear so, because they rotate in the plane of the sky \citep{montes21}. Such a motion, that cannot be observed, would bias the estimates of the mass downward. Our correlation indicates that is not likely. In such a case, the whole correlation would have to be interpreted such that the higher the surface brightness of the galaxy is, the more it rotates in the plane of the sky, which seems absurd.  On the other hand, some rotation would indeed help to align some galaxies more tightly with the correlation.

\begin{acknowledgements}
We thank the referee for a constructive report that helped to improve the quality of the manuscript. 
This research was supported by the Ministry of Science, Technological Development and Innovation of the Republic of Serbia under contract no. 451-03-33/2026-03/200002 with the Astronomical Observatory of Belgrade. 

\end{acknowledgements}

\bibliographystyle{aa}
\bibliography{literature,tables/lit2,tables/litm31,tables/bibliograpghyGannon,tables/litLVSo,tables/litLVIo}

\begin{appendix}
\section{Galaxy sample}
\label{app:data}

\subsection{Calculation of galaxy properties and their uncertainties}
For convenience of the potential followup works, we provide below tables of $m_{1/2}$, \rh, $M$, $D$, $x$, and $y$ for each object in our sample.  In this section, we describe the statistical method to estimate their most probable values of the target quantities and their uncertainties. We remind that in our work, the uncertainties were not used for finding the reported correlation. Only the uncertainties in $y$ were used for fitting $n$ in \equ{result}.

The data we used come from various sources. Different types of objects required slightly different approach of calculating the quantities of interest. The detailed way of calculation for each type of object is described in the next subsections. We however used a similar strategy to determine the uncertainties.
The calculation starts from some input quantities. The quantities of interest are their functions. To propagate the uncertainties of the input quantities into the quantities of interest, we generated $10^5$ random realizations of the input quantities according their expected probability distribution. Then, for each random realization, the standard formulas were used to get a realization of the quantity of interest. From the obtained distribution of the target quantity, we determined the 15.9, 50, and 84.1 percentiles to get the its expected value and the upper and lower 1-$\sigma$ uncertainty limits.

For the uncertainty distributions of the input quantities, we used either the normal or lognormal distribution, as indicated in the sections below. The peak value of the lognormal distribution chosen such that the mean of the distribution coincides with the measured value.  Whenever upper and lower uncertainties were given by the sources for some quantity, we averaged them, to get the width of the  normal or lognormal distribution.

In the published tables, we give either an upper and lower uncertainty limit, or just one uncertainty value.  The single uncertainty value means that the upper and lower limits are virtually identical.

\begin{table*}
\caption{Satellites of the MW.}          
\label{tab:mwsat} 
\centering            
\begin{tabular}{p{2.4cm}p{1.2cm}p{1.2cm}p{1.2cm}p{1.2cm}p{1.3cm}p{1.3cm}p{0.5cm}p{0.5cm}p{0.5cm}p{0.5cm}}
\hline\hline
Name & $\log_{10}\frac{m_{1/2}}{M_\sun}$ & \rh\newline[pc]  & $D$\newline[kpc] & $\sigma$\newline[km\,s$^{-1}$] & $\log_{10}x$ & $\log_{10}y$ & Ref.\newline$m_V$ & Ref.\newline$\re$ & Ref.\newline$\mu$  & Ref.\newline$\sigma$\\
\hline                                  
\csname @@input\endcsname
"tables/mwsat.txt"
\hline      
\end{tabular}
\tablefoot{The name is given in the same format as in the LVDB. The last four columns give references to the original source of the apparent $V$-band magnitude, angular effective radius, distance modulus, and velocity dispersion, respectively, according to the list below. \\
\textbf{References}: (1) \citet{2016MNRAS.463..712T}; (2) \citet{2023ApJ...950..167B}; (3) \citet{2018ApJ...860...66M}; (4) \citet{2006ApJ...653L.109D}; (5) \citet{2011ApJ...736..146K}; (6) \citet{2008ApJ...688..245W}; (7) \citet{2009MNRAS.397L..26C}; (8) \citet{2020ApJ...892...27M}; (9) \citet{2018ApJ...865....7C}; (10) \citet{2009ApJ...702L...9C}; (11) \citet{2008ApJ...674L..81K}; (12) \citet{2007ApJ...670..313S}; (13) \citet{2008ApJ...675L..73G}; (14) \citet{2015AJ....150...90K}; (15) \citet{2009AJ....137.3100W}; (16) \citet{2025arXiv250104772C}; (17) \citet{2021AJ....162..253M}; (18) \citet{2024ApJ...961..234H}; (19) \citet{2009ApJ...695L..83M}; (20) \citet{2024AJ....167..247B}; (21) \citet{2015MNRAS.448.2717W}; (22) \citet{2016ApJ...824L..14C}; (23) \citet{2021MNRAS.508.1064M}; (24) \citet{2017ApJ...838....8L}; (25) \citet{2021ApJ...920L..44C}; (26) \citet{2019ApJ...881..118W}; (27) \citet{2022ApJ...929..116O}; (28) \citet{2021ApJ...916...81C}; (29) \citet{2019MNRAS.490.2183M}; (30) \citet{2022ApJ...939...41C}; (31) \citet{2020ApJ...902..106M}; (32) \citet{2014PASP..126..616S}; (33) \citet{2008ApJ...675..201M}; (34) \citet{2005MNRAS.360..185B}; (35) \citet{2017ApJ...836..202S}; (36) \citet{2018ApJ...855...43M}; (37) \citet{2021ApJ...920...92J}; (38) \citet{2022ApJ...933..217R}; (39) \citet{2016ApJ...833...16K}; (40) \citet{2023ApJ...942..111C}; (41) \citet{2012ApJ...756...79S}; (42) \citet{2015ApJ...810...56K}; (43) \citet{2015MNRAS.454.1509M}; (44) \citet{2007ApJ...654..897B}; (45) \citet{2011ApJ...733...46S}; (46) \citet{2018A&A...609A..53C}; (47) \citet{2009ApJ...703..692L}; (48) \citet{2015ApJ...805..130K}; (49) \citet{2020ApJS..247...35V}; (50) \citet{2023AJ....165...55C}; (51) \citet{2020ApJ...892..137S}; (52) \citet{2013ApJ...767...62G}; (53) \citet{2012ApJ...752...42D}; (54) \citet{2024arXiv241012433G}; (55) \citet{2018AJ....156..257S}; (56) \citet{2006astro.ph..3486W}; (57) \citet{2011AJ....142..128W}. }
\end{table*}

\begin{table*}
\caption{Satellites of M\,31.}          
\label{tab:andsat} 
\centering            
\begin{tabular}{p{2.4cm}p{1.2cm}p{1.2cm}p{1.2cm}p{1.2cm}p{1.3cm}p{1.3cm}p{0.5cm}p{0.5cm}p{0.5cm}p{0.5cm}}
\hline\hline
Name & $\log_{10}\frac{m_{1/2}}{M_\sun}$ & \rh\newline[pc]  & $D$\newline[kpc] & $\sigma$\newline[km\,s$^{-1}$] & $\log_{10}x$ & $\log_{10}y$ & Ref.\newline$m_V$ & Ref.\newline$\re$ & Ref.\newline$\mu$  & Ref.\newline$\sigma$\\
\hline                                  
\csname @@input\endcsname
"tables/m31sat.txt"
\hline      
\end{tabular}
\tablefoot{The name is given in the same format as in the LVDB. The last four columns give references to the original source of the apparent $V$-band magnitude, angular effective radius, distance modulus, and velocity dispersion, respectively, according to the list below. \\
\textbf{References}: (1) \citet{2016ApJ...833..167M}; (2) \citet{2022ApJ...938..101S}; (3) \citet{2012ApJ...752...45T}; (4) \citet{2012ApJ...758..124H}; (5) \citet{2006MNRAS.365.1263M}; (6) \citet{2013ApJ...768..172C}; (7) \citet{2017ApJ...850..137M}; (8) \citet{2020MNRAS.491.3496C}; (9) \citet{2021MNRAS.505.5686C}; (10) \citet{2023MNRAS.521.3527C}; (11) \citet{2011ApJ...732...76R}; (12) \citet{2015ApJ...806..230S}; (13) \citet{2021MNRAS.503..176H}; (14) \citet{2013ApJ...768...50T}; (15) \citet{2023AJ....166..180R}; (16) \citet{2014ApJ...793L..14M}. }
\end{table*}

\begin{table*}
\caption{Local Volume satellite dwarfs.}          
\label{tab:lvs} 
\centering            
\begin{tabular}{p{2.4cm}p{1.2cm}p{1.2cm}p{1.2cm}p{1.3cm}p{1.3cm}p{0.5cm}p{0.5cm}p{0.5cm}p{0.5cm}}
\hline\hline
Name & $\log_{10}\frac{m_{1/2}}{M_\sun}$ & \rh\newline[pc]  & $\sigma$\newline[km\,s$^{-1}$] & $\log_{10}x$ & $\log_{10}y$ & Ref.\newline$m_V$ & Ref.\newline$\re$  & Ref.\newline$\sigma$\\
\hline                                  
\csname @@input\endcsname
"tables/LVSo.txt"
\hline      
\end{tabular}
\tablefoot{The name is given in the same format as in the LVDB. The last three columns give references to the original source of the apparent $V$-band magnitude, angular effective radius, and velocity dispersion, respectively, according to the list below. \\
\textbf{References}: (1) \citet{2018MNRAS.475.5085T}; (2) \citet{2018ApJ...857..145L}; (3) \citet{2024ApJ...967...72R}; (4) \citet{2015ApJ...811...62K}; (5) \citet{2018MNRAS.479.5343K}; (6) \citet{2018ApJ...863...25M}; (7) \citet{2015ApJ...808..108W}; (8) \citet{2016ApJ...833..167M}; (9) \citet{2013ApJ...768..172C}; (10) \citet{2019ApJ...884..128O}; (11) \citet{2023MNRAS.520.6312A}. }
\end{table*}

\begin{table*}
\caption{Local Volume isolated dwarfs.}          
\label{tab:lvi} 
\centering            
\begin{tabular}{p{2.4cm}p{1.2cm}p{1.2cm}p{1.2cm}p{1.3cm}p{1.3cm}p{0.5cm}p{0.5cm}p{0.5cm}p{0.5cm}}
\hline\hline
Name & $\log_{10}\frac{m_{1/2}}{M_\sun}$ & \rh\newline[pc]  & $\sigma$\newline[km\,s$^{-1}$] & $\log_{10}x$ & $\log_{10}y$ & Ref.\newline$m_V$ & Ref.\newline$\re$  & Ref.\newline$\sigma$\\
\hline                                  
\csname @@input\endcsname
"tables/LVIo.txt"
\hline      
\end{tabular}
\tablefoot{The name is given in the same format as in the LVDB. The last three columns give references to the original source of the apparent $V$-band magnitude, angular effective radius, and velocity dispersion, respectively, according to the list below.  \\
\textbf{References}: (1) \citet{2016ApJ...833..167M}; (2) \citet{2021MNRAS.503..176H}; (3) \citet{2024ApJ...972..180K}; (4) \citet{2012AJ....144....4M}; (5) \citet{2014MNRAS.439.1015K}; (6) \citet{2020A&A...635A.152T}. }
\end{table*}

\begin{table*}
\caption{Coma Cluster UDGs.}          
\label{tab:udgcoma} 
\centering            
\begin{tabular}{p{3.4cm}p{1.2cm}p{1.2cm}p{1.2cm}p{1.2cm}p{1.2cm}p{1.3cm}p{1.3cm}}
\hline\hline
Name & $\log_{10}\frac{m_{1/2}}{M_\sun}$ & \rh\newline[kpc]  & $D$\newline[Mpc] & $\sigma$\newline[km\,s$^{-1}$]& $\log_{10}\frac{M}{M_\sun}$ & $\log_{10}x$ & $\log_{10}y$ \\
\hline                                  
\csname @@input\endcsname
"tables/UDGComa.txt"
\hline      
\end{tabular}
\end{table*}

\begin{table*}
\caption{dark-matter deficient UDGs.}          
\label{tab:udgdmfree} 
\centering            
\begin{tabular}{p{1.4cm}p{1.2cm}p{1.2cm}p{1.2cm}p{1.3cm}p{1.3cm}}
\hline\hline
Name & $\log_{10}\frac{m_{1/2}}{M_\sun}$ & \rh\newline[kpc]  & $\sigma$\newline[km\,s$^{-1}$]&  $\log_{10}x$ & $\log_{10}y$ \\
\hline                                  
\csname @@input\endcsname
"tables/UDGDMfree.txt"
\hline      
\end{tabular}
\end{table*}

\begin{table*}
\caption{UDGs of Gannon catalog.}          
\label{tab:gudg} 
\centering            
\begin{tabular}{p{2.7cm}p{1.2cm}p{1.2cm}p{1.2cm}p{1.3cm}p{1.3cm}p{2.5cm}}
\hline\hline
Name & $\log_{10}\frac{m_{1/2}}{M_\sun}$ & \rh\newline[kpc]  & $\sigma$\newline[km\,s$^{-1}$] & $\log_{10}x$ & $\log_{10}y$ & Ref. \\
\hline                                  
\csname @@input\endcsname
"tables/GUDG.txt"
\hline      
\end{tabular}
\tablefoot{The name is given in the format used in the the catalog of \citet{gannon24}. The last columns gives the references to the original sources. \\
\textbf{References}: (1) \citet{MartinezDelgado2016}; (2) \citet{MartinNavarro2019}; (3) \citet{Janssens2022}; (4) \citet{Iodice2020}; (5) \citet{Iodice2023}; (6) \citet{Buttitta2025}; (7) \citet{Forbes2019}; (8) \citet{Muller2020}; (9) \citet{Muller2021}; (10) \citet{Forbes2021}; (11) \citet{Danieli2022}; (12) \citet{FerreMateu2023}; (13) \citet{Lim2020}; (14) \citet{Toloba2023}; (15) \citet{Gannon2022}; (16) \citet{Janssens2024}; (17) \citet{Gannon2021}; (18) \citet{Beasley2016}; (19) \citet{Gannon2020}; (20) \citet{Toloba2018}; (21) \citet{Yagi2016}; (22) \citet{Alabi2018}; (23) \citet{Chilingarian2019}; (24) \citet{FerreMateu2018}; (25) \citet{vanDokkum2017}; (26) \citet{Lim2018}; (27) \citet{Gannon2023}. }
\end{table*}

\subsection{Dwarf spheroidals}
For dwarf spheroidals, the data were taken from the Local Volume Database\footnote{\url{https://github.com/apace7/local_volume_database}} (LVDB, \citealp{pace25}). This is an online  continually updated database of many parameters of nearby galaxies and star clusters collected from the literature. We used the version from 8 May 2025\footnote{A link to a  backup at Zenodo of the used table will be placed here after the acceptance.}. This database was separated into several subsets, following the headings of the subsection below. From all subsets, we excluded the galaxies that have any detection of gas and those for which we could not calculate the quantity of interest, that is $\aobs/\abar$, because some of the quantities necessary for its calculation, i.e. the effective radius, magnitude, and velocity dispersion, are missing in the database. 

For all dwarf spheroidals, we assumed the stellar mass-to-light ratio in the $V$-band of two, corresponding to an old stellar population.  A normal distributions in the distance modulus, apparent magnitude, angular effective radius, and velocity dispersion were assumed as given in the LVDB. 

\subsubsection{Milky Way satellites}
For the satellites of the MW, we further excluded Antilia\,II and Crater\,II, because they seem to be in an advanced stage of tidal disruption \citep{ji21}. 
Our final MW sample consists of the 30 objects listed in \tab{mwsat}.

The galactocentric distances were calculated from the distance moduli and sky coordinates in the LVDB and the distance of the MW center from the Sun  8.27\,kpc \citep{gaia23}. We assumed the baryonic mass of the MW of $6\,\times\,10^{10}\,M_\sun$ \citep{zhao13}. 

The resolved velocity dispersion profiles for the MW satellites come from \citet{walker07}. For the modelling, we assumed a Sérsic index of one.

\subsubsection{M\,31 satellites}
For the satellites of M\,31, we additionally excluded the M\,32 galaxy, because it is not a dwarf spheroidal. 
The 27 used objects are listed in \tab{andsat}

The distances from M\,31 center were calculated from their sky coordinates, their heliocentric distances, and the heliocentric distance of M\,31, which was assumed to be $761\pm11\,$kpc \citep{li21}. A gaussian error distribution was assumed for the distance of M\,31. We assumed the baryonic mass of M\,31 of $12\,\times\,10^{10}\,M_\sun$ \citep{zhao13}.

\subsubsection{Local Volume dwarfs}
The LVDB lists also data for nearby dwarf galaxies that are not satellites of the MW or M\,31. Sime of then are isolated, some are satellites of other galaxies. We excluded from both groups the galaxies that are not classified as dwarf spheroidals.

The satellite group includes  Carina\,II,  Carina\,III, Horologium\,I, Hydrus\,I and Reticulum\,II as the satellites of the Large Magellanic Cloud. Next, it includes Andromeda\,XXII a satellite of M\,33, and KDG\,64, which is a satellite of M\,81. In total, the sample of Local Volume satellites amounts seven objects. They are listed in \tab{lvs}. The sample of isotated Local Volume dwarfs contains the three objects listed in  \tab{lvi}. 

\subsection{UDGs}
\subsubsection{Coma UDGs}
The sample of the UDGs in the Coma Cluster is adopted from \citet{freundlich22}. It is a compilation of ten objects from the literature. The authors also provided the most probable three dimensional distances from the center of Coma and the radial baryonic mass profile of the cluster itself. Almost all, namely eight, data on these objects come from \citet{chilingarian19}, where their velocity dispersion was measured using a high-resolution spectrograph such that the spectral line profiles are properly sampled. For them, also the stellar mass-to-light ratios derived from the spectra were available. All of this is makes the Coma UDGs more reliable than the other UDG datasets, which is why we used these UDGs in our initial search of the prescription for predicting the gravitational fields from the baryons. For the Coma UDGs, we used as the mass $M$ of the host the baryonic mass of the Coma Cluster cumulated below the estimated cluster-centric distance of each given dwarf \citet{freundlich22}. 

The stellar masses were calculated from the luminosities and the mass-to-light ratios given in \citet{freundlich22}. The original magnitude, effective radius and velocity dispersion come from \citet{vandokkum17b} for of DFX1, and from \citet{vandokkum19} for DF44. For the remaining Coma UDGs, the radii, magnitudes, velocity dispersions and mass-to-light ratios come from \citet{chilingarian19}. The other data  originate from \citet{freundlich22} themselves.

For the velocity dispersion and mass-to-light ratio, we used normal distributions, with the uncertainties from \citet{freundlich22}. For the other quantities, we did not have any information about the observational uncertainties. We instead used an educated guess. For the 3D distance $D$, we assumed a lognormal distribution of uncertainty given with a standard deviation of $0.5\log_{10}(D/D_p)$, where $D_p$ is the projected distance of the galaxy from the cluster center. A lognormal distribution with a $\log_{10}1.1$ uncertainty was assumed for the effective radii. The uncertainty for $M$ stems from the uncertainty in the 3D distance, and we added another $\log_{10}1.5$ of lognormal scatter to account for the uncertainty of the distribution of baryonic mass in the cluster.

\subsubsection{Dark-matter deficient UDGs}
We are aware of four gasless galaxies that were specifically advertised as dark-matter deficient in the literature: the DF\,2, DF\,4 \mb{and DF\,9} UDGs in the vicinity of the NGC\,1052 galaxy, and FCC\,224 near the Fornax Cluster. For the DF\,2 and DF\,4 galaxies the luminosities and effective radii were taken from \citet{muller19}. For DF\,2, we used the observed velocity dispersion from \citet{emsellem19}, and for DF\,4 from \citet{vandokkum18}. For FCC\,224, all parameters were taken from \citet{buzzo25} \mb{and for DF\,9 from \citep{keim26}. }

For DF\,2, DF\,4 \mb{and DF\,9}, the stellar masses were calculated from the luminosities and the mass-to-light ratios. A stellar mass-to-light ratio of two was assumed with a lognormal scatter of $\log_{10}1.5$. 
For FCC\,224, the stellar mass and its uncertainty were taken directly from \citet{buzzo25}. The uncertainty was assumed to have a normal distribution.
For all dark-matter deficient galaxies, a lognormal distribution with a $\log_{10}1.1$ uncertainty was assumed for the effective radii. A normal distribution of velocity dispersion was assumed with widths according to the measurements. All final parameters of this data group are given in \tab{udgdmfree}.

\subsubsection{UDG catalog by \citet{gannon24}}
The catalog by \citet{gannon24}\footnote{\url{https://github.com/gannonjs/Published_Data/tree/main/UDG_Spectroscopic_Data}} is a continually updated online  database of parameters of UDGs complied from the literature. We used the version from 27 August 2025\footnote{A link to a  backup at Zenodo of the used table will be placed here after the acceptance.}. Just as for the LVDB, we excluded galaxies for which the catalog does not list the effective radius, luminosity, or velocity dispersion. We also removed the galaxies included already in the previous datasets. 
We ended up with the 23 objects in \tab{gudg}.

A stellar mass-to-light ratio of two was assumed with a lognormal scatter of $\log_{10}1.5$. A lognormal distribution with a $log_{10}1.2$ uncertainty was assumed for the effective radii.  The uncertainty of $\sigma$ was assumed to have a normal distribution with a width as listed in the database.

In \sect{resolved}, we modeled the resolved velocity dispersion profiles of the UDGs DF\,44 and NGC\,5846\_UDG1. For DF\,44, the profile comes from \citet{vandokkum19}. We adopted the parameters of the galaxy from the paper \citet{bil19c}. The velocity dispersion of NGC\,5846\_UDG1 was taken from \citet{haacke25}, where it was derived from globular clusters. For its modeling, the parameters of the galaxy were taken from the same paper.

\section{Confidence intervals for Spearman correlation}
\label{app:spearmann}
In \sect{finding} we calculated the Spearman correlation coefficient on a grid of parameters  $p_2$ and $p_3$. Given that some pair of  $p_{2\mathrm{,max}}$  and $p_{3\mathrm{,max}}$  gives the strongest correlation, we needed to determine for which combinations of  $p_2$ and $p_3$ the correlation is statistically weaker. To this end, we made use of the fact at  $p_{2\mathrm{,max}}$  and $p_{3\mathrm{,max}}$, the correlation has the shape of a power law $y=nx^\alpha$. In the log-log space, we fitted the parameters and noted the residuals from $y$, that is $d\log_{10}y_i  =\log_{10}y_i-\log_{10}y(x_i)$, and the correlation coefficient $\rho_\mathrm{max}$. Then we 1000 times randomly redistributed the residuals around the fit, and calculated the Spearman correlation coefficient $\rho$. We calculated the 1-$\sigma$ width of the distribution of $\rho$ as $w = \sqrt{\sum(\rho - \rho_\mathrm{max})^2 /1000}$. With this definition of $w$, we ensure that differences between the Spearman coefficients of two models can be accounted to the noise in the data at the 1-$\sigma$ confidence if the values of their $\rho$ differ by less than $w$. Then, on the whole map of $\rho$ on grid of $p_2$ and $p_3$, we could decide which locations differ from the best location by less than a given confidence limit.

\end{appendix}
\label{LastPage} 
\end{document}